# Many-body theories for negative kinetic energy systems

Huai-Yu Wang[a]

*Department of Physics,Tsinghua University, Beijing, 100084 China*

**Abstract:** In the author's previous works, it is derived from the Dirac equation that particles can have negative kinetic energy (NKE) solutions, and they should be treated on an equal footing as the positive kinetic energy (PKE) solutions. More than one NKE particles can make up a stable system by means of interactions between them and such a system has necessarily negative temperature. Thus, many-body theories for NKE systems are desirable. In this work, the many-body theories for NKE systems are presented. They are Thomas-Fermi method, Hohenberg-Kohn theorem, Khon-Sham self-consistent equations, and Hartree-Fock self-consistent equations. They are established imitating the theories for PKE systems. In each theory, the formalism of both zero temperature and finite negative temperature are given. In order to verify that tunneling electrons are of NKE and real momentum, an experiment scenario is suggested that lets PKE electrons collide with tunneling electrons.

**Keywords:** Negative kinetic energy Schrödinger equation, negative kinetic energy system, negative temperature, many-body theory, Thomas-Fermi method, Hohenberg-Kohn theorem, Kohn-Sham equation, Hartree-Fock self-consistent equation

[a] wanghuaiyu@mail.tsinghua.edu.cn

Résumé: Dans les travaux antérieurs de l'auteur, il découle de l'équation de Dirac que les particules peuvent avoir des solutions d'énergie cinétique négative (NKE), et elles doivent être traitées sur un pied d'égalité avec les solutions d'énergie cinétique positive (PKE). Plusieurs particules NKE peuvent constituer un système stable au moyen d'interactions entre elles et un tel système a nécessairement une température négative. Ainsi, les théories à plusieurs corps pour les systèmes NKE sont souhaitables. Dans ce travail, les théories à plusieurs corps pour les systèmes NKE sont présentées. Ce sont la méthode de Thomas-Fermi, le théorème de Hohenberg-Kohn, les équations auto-cohérentes de Khon-Sham et les équations auto-cohérentes de Hartree-Fock. Ils sont établis en imitant les théories des systèmes PKE. Dans chaque théorie, le formalisme de la température nulle et de la température négative finie est donné. Afin de vérifier que les électrons tunnel sont de NKE et d'impulsion réelle, un scénario expérimental est proposé qui laisse les électrons PKE entrer en collision avec des électrons tunnel.

# I. INTRODUCTION

The Schrödinger equation in quantum mechanics (QM) was proposed in the case that a particle's energy $E$ is greater than potential $V$, i.e.,

$$i\hbar \frac{\partial}{\partial t}\psi = (-\frac{\hbar^2}{2m}\nabla^2 + V)\psi, \quad E > V. \tag{1}$$

The constrain $E > V$ means that a particle is of positive kinetic energy (PKE). People applied the Schrödinger equation to region $E < V$ and found that there were solutions, too. That revealed the famous effect of potential barrier penetration. However, the author first noticed[1] that whether this application was legitimate or not has never been experimentally quanlitatively verified nor theoretically derived. The fact is that there has been no experiment detecting particles inside a potential barrier, i.e., inside a region where $E < V$. Nevertheless, the author thought that it was possible to derive a QM equation of a particle in the region $E < V$ for low-momentum motion.

It is well known that in classical mechanics, Newtonian mechanics applies to the low-momentum motion of bodies. Einstein developed relativistic mechanics. When the low-momentum approximation of the relativistic mechanics is made, one obtains the formulas of the Newtonian mechanics.

In QM, the case is similar. When low-momentum approximation of the relativistic QM equation is made, the Schrödinger equation is obtained. It is indeed so. We take the Dirac equation for spin-1/2 particles as an example.

$$i\hbar \frac{\partial}{\partial t}\Psi = (c\boldsymbol{\alpha} \cdot \boldsymbol{p} + mc^2 \beta + V)\Psi. \tag{2}$$

We take the following transformation for wave function $\Psi$.

$$\Psi = \psi e^{-imc^2 t/\hbar}. \tag{3}$$

When Eq. (3) is substituted into (2) and low-momentum approximation is made, the Schrödinger equation (1) is obtained. However, there is actually a transformation other than (3):

$$\Psi = \psi_{(-)} e^{imc^2 t/\hbar}. \tag{4}$$

Substituting Eq. (4) into (2) and making low-momentum approximation, we obtain

$$i\hbar \frac{\partial}{\partial t}\psi_{(-)} = (\frac{\hbar^2}{2m}\nabla^2 - V)\psi_{(-)}. \tag{5}$$

This is calld negative kinetic energy (NKE) Schrödinger equation.[1]

The author thinks that the two transformations (3) and (4) are of equal status. Consequently, the Schrödinger equation and NKE Schrödinger equation are also of equal status. Yet in the Schrödinger equation, the kinetic term has its classical correspondence, whilst in classical mechanics, kinetic energy is always nonnegative.

However, in Eq. (5), the kinetic energy term has a minus sign compared to that in (1), which means that the kinetic energy is negative. That is why the author called it NKE Schrödinger equation. Actually, when establishing the waved equation of QM, Schrödinger himself, in the fourth one of his series papers,[2] achieved Eqs. (1) and (5) simultaneously, and he thought the two equations were mathematically equivalent. Nevertheless, he adopted Eq. (1). Since then, Eq. (5) has never been mentioned.

The Dirac equation itself has both PKE and NKE solutions. It is believed that they have equivalent status. Making use of this equivalence, we figured out the famous Klein's paradox when a free Dirac particle encountered a potential barrier.[3] Since the Dirac equation is correct, its two low-momentum approximations (1) and (5) are naturally correct either. We believe that there is a PKE-NKE symmetry in the universe, and this is a great symmetry. This symmetry has at least two special characteristics. One is that it has not been known what group the two elements PKE and NKE belong to and which operation can turn one to another. The other is that up to date, no one has found any matter with NKE. The concept of NKE is against common sense.

The author's viewpoint is that NKE objects are dark ones. That is why people have never met them before. Since the NKE matter is dark, it is desirable to find out it. We have suggested experiments to verify the existence of NKE electrons.[1] The scenario was to let photons impinge tunneling electrons. In Appendix, we suggest another scenario: PKE electrons can be used to collide with tunneling electrons.

The NKE-PKE symmetry in our universe means that they should be treated on an equal footing. Thus, all the field concerning the PKE ought to be revisited in the aspect of the NKE.[1,3–12] In Appendix D of [1], 13 points were listed which were topics to be dealt with in the following work. The point 1 was solved in [9], point 2 solved by [3], point 4 by [4], points 6 and 7 by [5], point 9 by [11], and points 5 and 13 by [12]. The present paper intends to deal with point 10. The fundamental equations of QM should be in symmetric forms with respect to the PKE and NKE.[6] The virial theorem also embodied the PKE-NKE symmetry.[7] Some one-dimensional problems in QM are resolved by the combination of the Schrödinger equation and NKE Schrödinger equation.[8] In the expressions of one-body Green's function, the NKE solutions should also be contained.[10]

It was found that NKE matter produced negative pressure.[4,5] In [12], a theory of dark energy that matched dark matter theory was given. The dark energy also produced negative pressure. A current popular view is that the accelerating expansion of the universe is due to negative pressure. Our research provided the sources of the negative pressure.

More than one NKE particles can compose a system by means of interactions between them, just as PKE particles do. The author has researched the simplest cases, the two-body systems[5] and the simplest two three-body systems.[11] We intend now to present the many-body theories for NKE systems.

Here we clarify the concepts of the PKE and NKE systems. A PKE system means that all the particles composing the system are of PKE, and electrons and nuclei have respectively negative and positive charges. There can be attractive and repulsive Coulomb interactions. A NKE system means that all the particles inside are of NKE and they all have the same kind of charge. Thus, the total charge of the system is not zero. It is possible that a system has a nonzero charge, for instance, a black hole with a charge.[13] Besides, it is possible that in a system, some particles are of PKE and others are of NKE, called a combo system. In this work we do not touch combo systems.

Since we think a NKE particle is a dark one, the NKE systems are called dark systems. The NKE matter and dark matter are regarded as synonym.

Up to now, all many-body theories have been developed for PKE systems. Now that NPE and PKE should be treated on an equal footing, for each many-body theory of PKE systems, there should be corresponding theory for NKE systems. We can simply mimic the theories of PKE systems[14] to establish the corresponding theories of NKE systems. In doing so, we should keep in mind that NKE systems are of negative temperature (NT). Therefore, for a NKE system, the higher the energy level the stabler the state, and the highest energy level is the ground state. The NKE systems observe energy maximum principle.[4]

We are going to present the many-body theories of NKE systems imitating those of PKE systems. Section II presents Thomas-Fermi method. Section III proves Hohenberg-Kohn theorem. Based on this theorem, Khon-Sham self-consistent equations are established in Section IV. Section V shows Hartree-Fock self-consistent equations. In each case, both zero-temperature and finite-negative-temperature theories are given. Section VI is a brief summary. In Appendix, experiments are suggested that let PKE and NKE electrons collide.

## II. THOMAS-FERMI METHOD

### A. Non-Interactive systems

We assume that a NKE system is composed of a nucleus and electrons, both carrying the same kind of electric charge, denoted by $q$, e.g., the nucleus also has negative charge.

We discuss the case of zero temperature. The nucleus is set to be static at the origin. The density of the NKE electron is $\rho_{(-)}(\boldsymbol{r})$. The electrostatic potential $\phi_{(-)}(\boldsymbol{r})$ should obey equation

$$\nabla^2 \phi_{(-)}(\boldsymbol{r}) = q\rho_{(-)}(\boldsymbol{r})/\varepsilon_0 . \tag{6a}$$

If $\rho_{(-)}(\boldsymbol{r})$ is of spherical symmetry with respect to the origin, either is the $\phi_{(-)}(\boldsymbol{r})$. Thus, Eq. (6a) can be recast into

$$\frac{1}{r^2}\frac{\mathrm{d}}{\mathrm{d}r}(r^2 \frac{\mathrm{d}}{\mathrm{d}r}\phi_{(-)}) = \frac{1}{r}\frac{\mathrm{d}^2(r\phi_{(-)})}{\mathrm{d}r^2} = \frac{q}{\varepsilon_0}\rho_{(-)}(r) . \tag{6b}$$

For the NKE system considered, $\phi_{(-)}(r)$ has to meet the following boundary

conditions:

$$\phi_{(-)}(r) = \frac{Zq}{4\pi\varepsilon_0 r}, r \to 0, \tag{7}$$

where Z is the charge number of the nucleus, and

$$\phi_{(-)}(r) = \frac{zq}{4\pi\varepsilon_0 r}, r > R, \tag{8}$$

where z is the charge number of the "ion". The dark $H^-$ discussed in [11] can be regarded as an examples. The R in (8) describes the size of the atom and its value is to be determined. At $r = R$, $\phi_{(-)}(r)$ and its first derivative should be continuous. Consequently, the boundary conditions are

$$\phi_{(-)}(R) = \frac{zq}{4\pi\varepsilon_0 R} \tag{9a}$$

and

$$(\frac{d\phi_{(-)}}{dr})_R = -\frac{zq}{4\pi\varepsilon_0 R^2}. \tag{9b}$$

Equation (6) is a relation between $\phi_{(-)}(r)$ and $\rho_{(-)}(r)$. There is another relation between them. At distance r, the electrons' energy and kinetic energy has a relation

$$E_{(-)} = -p^2/2m + q\phi_{(-)}(r). \tag{10}$$

Since we are discussing bounded electrons, their total energy should be smaller than their potential energy at boundary $r = R$. Suppose that the largest momentum is $p_{max}$. Then,

$$-p_{max}^2/2m + q\phi_{(-)}(r) = q\phi_{(-)}(R) \tag{11a}$$

or

$$p_{max}^2/2m = q[\phi_{(-)}(r) - \phi_{(-)}(R)]. \tag{11b}$$

The largest momentum $p_{max}$ is Fermi momentum $p_f$. In a Fermi sphere, electron density is

$$\rho_{(-)} = \frac{k_f^3}{3\pi^2} = \frac{p_f^3}{3\pi^2\hbar^3}. \tag{12}$$

We think that Eq. (12) stands at any point, i.e., the left and right hand sides of (12) are functions of position *r*. Equating (12) and (11b) lead to

$$\rho_{(-)}(r) = \frac{1}{3\pi^2 \hbar^3}[2mq(\phi_{(-)}(r) - \phi_{(-)}(R))]^{3/2}. \tag{13}$$

Thus, we have two relations between potential $\phi_{(-)}(r)$ and density $\rho_{(-)}(r)$, Eqs. (6) and (13). The latter is substituted into the former to get

$$\frac{1}{r}\frac{d^2(r\phi_{(-)})}{dr^2} = \frac{4e}{3\pi \hbar^3 \varepsilon_0}[2mq(\phi_{(-)}(r) - \phi_{(-)}(R))]^{3/2}. \tag{14}$$

This is the equation for potential $\phi_{(-)}(r)$, and in principle it is solvable. Obviously, it is difficult to obtain analytical solutions, so, numerical computation has to be carried out.

It is possible to reform Eq. (14) to be a more concise form. We define a dimensionless function

$$\varphi_{(-)}(r) = \frac{r}{Ze}(\phi_{(-)}(r) - \phi_{(-)}(R)). \tag{15a}$$

This means that

$$r\phi_{(-)}(r) = Zq\varphi_{(-)}(r) + r\phi_{(-)}(R)). \tag{15b}$$

Taking second derivatives with respect to $r$ and then substituting it into (14), one gets

$$\frac{d^2 \varphi_{(-)}(r)}{dr^2} = \frac{4q^3}{3Z\pi \hbar^3 \varepsilon_0}\frac{(2Zm)^{3/2}}{\sqrt{r}}\varphi_{(-)}^{3/2}(r). \tag{16}$$

Substituting (7) into (15a), one obtains the boundary conditions that $\varphi_{(-)}(r)$ satisfies are

$$\varphi_{(-)}(r = 0) = \frac{1}{4\pi \varepsilon_0} \tag{17}$$

and

$$\varphi_{(-)}(R) = 0. \tag{18}$$

It follows from Eq. (9) that the condition that $\varphi_{(-)}(r)$ is smooth at the boundary is that

$$(\frac{d\varphi_{(-)}}{dr})_R = \frac{1}{Zq}\frac{d}{dr}[r(\phi_{(-)}(r) - \phi_{(-)}(R))]_{r=R} = -\frac{z}{4\pi \varepsilon_0 ZR}. \tag{19}$$

Because the interaction between electrons is not taken into account, electrons have merely NKE and potential energies. Both the terms are just contrary to those of PKE systems. Therefore, the above consideration is formally just the copy of Thomas-Fermi method of PKE systems.

**B. Interactive systems**

Now, the interaction between electrons is taken into account.

At any point in space, electron density and momentum have relationship (12). The chemical potential is a constant. At every point, the sum of the kinetic and potential energies should be the chemical potential. Since now kinetic energy is negative, chemical potential is as well.[4]

$$\mu_{(-)} = -p_f^2(r)/2m + V(r). \tag{20}$$

In the present case, the nucleus and electrons carry the same kind of charge. Therefore, the external potential $V(r)$ is positive. By use of Eq. (12), the right hand side of (20) is expressed by density:

$$\mu_{(-)} = -\hbar^2(3\pi^2)^{2/3}\rho_{(-)}^{2/3}(r)/2m + V(r). \tag{21}$$

This is the Thomas-Fermi equation. Because kinetic energy is negative, in the region $\mu_{(-)} - V(r) > 0$, the density is zero, $\rho_{(-)}(r) = 0$. Hereafter, we may omit the subscript (−) of the density $\rho$ in the case of unambiguity.

It is well known that for a non-interactive Fermi gas, the average kinetic energy is 3/5 of Fermi energy. Hence, the total kinetic energy $K_0$ of electron gas can be written as

$$K_0 = -\frac{3N}{5}\frac{1}{2m}p_f^2(r). \tag{22}$$

where $N$ is total number of electrons. The kinetic energy in a unit volume $t_0$ is

$$k_0 = \frac{K_0}{V} = -\frac{3}{5}\frac{p_f^2}{2m}\rho_0 = -c_k \rho_0^{5/3}, \tag{23}$$

where $c_k$ is

$$c_k = \frac{3\hbar^2}{10m}(3\pi^2)^{2/3}. \tag{24}$$

The two sides of (23) are also written as the functions of coordinates,

$$k(r) = -c_k \rho^{5/3}(r). \tag{25}$$

The ground state energy is expressed by

$$E = -c_k \int dr \rho^{5/3}(r) + \int dr \rho(r) V_N(r) + \frac{e^2}{2}\int dr dr' \frac{\rho(r)\rho(r')}{|r-r'|}, \tag{26}$$

where $V_N(r)$ is the external potential the electrons are subject to, e.g., that from nucleus. Variational principle is utilized:

$$\delta(E - \mu N)/\delta\rho = 0. \tag{27}$$

The chemical $\mu$ plays a role of Lagrange multiplier. The constraint condition is normalization,

$$\int d\mathbf{r}\rho(\mathbf{r}) = N. \tag{28}$$

The variation result is

$$\mu = -\frac{5}{3}c_k\rho^{2/3}(\mathbf{r}) + V_N(\mathbf{r}) + e^2 \int d\mathbf{r}' \frac{\rho(\mathbf{r}')}{|\mathbf{r}-\mathbf{r}'|}. \tag{29}$$

By variation (27), the energy maximum of the system is obtained. This obeys the energy maximum principle of NKE systems.[4]

We do not consider exchange interaction. It is anticipated that if the exchange interaction is included, the total energy will be raised, since the higher the energy, the more stable the system.

We turn to the case of finite NT.

Both Eqs. (12) and (23) are expressions for zero temperature. They are not valid at finite NT. For PKE systems, the generalization from zero to finite temperature has been discussed.[15–17] We follow this approach.

At finite NT, electron number in phase space is

$$\frac{2}{(2\pi\hbar)^3} \frac{4\pi p^2 dp d\tau}{\exp[\beta(-p^2/2m + V(\mathbf{r}) - \mu)] + 1}, \tag{30}$$

where $V(\mathbf{r})$ is potential including that generated by the nucleus and those from other electrons. Chemical potential $\mu$ is now a function of NT. We denote

$$\beta = \frac{1}{k_B T}, T \leq 0, \tag{31}$$

where $k_B$ is Boltzmann constatnt and is dropped hereafter. At thermo equilibrium, the electron number in unit volume is

$$\rho(\mathbf{r}) = \int_0^\infty dp \frac{2}{(2\pi\hbar)^3} \frac{4\pi p^2}{\exp[\beta(-p^2/2m + V(\mathbf{r}) - \mu)] + 1}. \tag{32}$$

In this case, the particle density in Eqs. (21)-(29) should use the expression (32). At each fixed NT, the density is solved by these equations self-consistently. When NT goes to zero, Fermi distribution approaches step function, and Eq. (32) degrades to (12).

### III. HOHENBERG-KOHN THEOREM

For PKE electron systems, there is a Hohenberg-Kohn theorem.[18] For NKE electron systems there can also be such a theorem.

#### A. Proof of the Hohenberg-Kohn theorem of NKE systems

Suppose that a system comprises $N$ NKE electrons. We consider its ground state, the energy of which is denoted by $E$. It is assumed that the ground state wave function $\Phi_G(\boldsymbol{r}, \boldsymbol{r}_2, \cdots, \boldsymbol{r}_N)$ has been normalized. The density of the system is

$$\rho(\boldsymbol{r}) = N \int d\boldsymbol{r}_2 \cdots d\boldsymbol{r}_N |\Phi_G(\boldsymbol{r}, \boldsymbol{r}_2, \cdots, \boldsymbol{r}_N)|^2. \tag{33}$$

The density is not negative,

$$\rho(\boldsymbol{r}) \geq 0. \tag{34}$$

Whatever happens, the density always observes normalization condition,

$$N = \int d\boldsymbol{r} \rho(\boldsymbol{r}). \tag{35}$$

Every electron may be subject to an external potential, such as nuclei.

$$V = \sum_{i=1}^{N} v(\boldsymbol{r}_i). \tag{36}$$

This potential energy can be rewritten by field operators:

$$V = \int d\boldsymbol{r} v(\boldsymbol{r}) \psi^+(\boldsymbol{r}) \psi(\boldsymbol{r}). \tag{37}$$

The density of the ground state is

$$\rho(\boldsymbol{r}) = \langle G | \psi^+(\boldsymbol{r}) \psi(\boldsymbol{r}) | G \rangle. \tag{38}$$

Under a potential $v(\boldsymbol{r})$, Hamiltonian is $H$ and ground state energy is $E = \langle G | H | G \rangle$.

In principle, potential can be added an arbitrary constant independent of coordinates. We do not consider the constant.

We are now at the stage to prove Hohenberg-Kohn theorem of NKE systems. As a matter of fact, since the ground state energy of a NKE system reaches its maximum, one merely needs to replace the less than sign in the theorem for PKE systems by greater than sign.

First, we prove that $v(\boldsymbol{r})$ is the only functional of density $\rho(\boldsymbol{r})$. This means that as soon as $\rho(\boldsymbol{r})$ is determined, there exists unique $v(\boldsymbol{r})$, by which the NKE Schrödinger equation containing the potential (37) is solved to get the density in (33), is just the assumed one.

Assume that there is another $v'(\boldsymbol{r})$, by which a corresponding Hamiltonian is $H'$. The corresponding ground state is denoted by $|G'\rangle$ and its energy is $E' = \langle G' | H' | G' \rangle$. We have the following inequality.

$$E' > \langle G | H' | G \rangle = \langle G | H + V' - V | G \rangle$$
$$= E + \langle G | V' - V | G \rangle = E + \int d\mathbf{r}[v'(\mathbf{r}) - v(\mathbf{r})]\rho(\mathbf{r}). \quad (39)$$

Suppose that $v'(\mathbf{r})$ and $v(\mathbf{r})$ result in identical $\rho(\mathbf{r})$. By the same reasoning, it can be obtained that

$$E > E' + \int d\mathbf{r}[v(\mathbf{r}) - v'(\mathbf{r})]\rho(\mathbf{r}). \quad (40)$$

The sum of the two equations leads to $E + E' < E + E'$, which is unreasonable. The conclusion is that $v(\mathbf{r})$ is unique.

Now the interaction $U$ between electrons is added. The free energy functional is

$$F[\rho(\mathbf{r})] = \langle G | K + U | G \rangle. \quad (41)$$

Ground state energy is

$$E_G[\rho] = F[\rho(\mathbf{r})] + \int d\mathbf{r} v(\mathbf{r})\rho(\mathbf{r}). \quad (42)$$

We intend to prove that as $\rho(\mathbf{r})$ reaches that of the real ground state, the $E_G[\rho]$ reaches its maximum. The process of the proof is as follows. Assume that $\Phi$ is correct ground state wave function, and $\Phi'$ is another one solved by NKE Schrödinger equation containing $v'(\mathbf{r})$ and corresponding density is $\rho'(\mathbf{r})$. Then, we have

$$E_G[\Phi'] = \langle G' | K + U | G' \rangle + \langle G' | V | G' \rangle = F[\rho'] + \int d\mathbf{r} v(\mathbf{r})\rho'(\mathbf{r})$$
$$< E_G[\Phi] = F[\rho(\mathbf{r})] + \int d\mathbf{r} v(\mathbf{r})\rho(\mathbf{r}). \quad (43)$$

Thus, it is seen that $E_G[\rho(\mathbf{r})]$ is the maximum among the densities corresponding to all possible $v'(\mathbf{r})$.

The Hohenberg-Kohn theorem for NKE systems comprises two points as follows.
(1) The ground state energy of an identical spinless Fermion system is the unique functional of density. (2) When the density subject to normalization condition reaches the correct value, this functional reaches its maximum.

Please note here that $F[\rho(\mathbf{r})]$ has to be $v(\mathbf{r})$-representative.[19,20]

## B. Inclusion of spin and relativistic motion

For PKE systems, Hohenberg-Kohn theorem has been extended to the cases having spin.[21–24] The treatment of NKE systems imitates that.

The inclusion of spin has to take into account a spin density $s(\mathbf{r})$ in the system besides the particle density $\rho(\mathbf{r})$. The latter is subject to an external potential $v(\mathbf{r})$ and the former to an external magnetic field $B(r)$.

In order to deal with this problem, it had better to start from relativistic motion, which includes spin. In this way, both spins and relativistic motion are considered. Now we inspect four-electric current $j_\mu(x)$ and four-potential $A_\mu(x)$, where $x$ represents four-coordinates.

In Fock space, the Dirac equation of electrodynamics is

$$(i\hbar \frac{\partial}{\partial t} - H)\psi = 0. \tag{44}$$

Hamiltonian is composed of four parts:

$$H = H_0 + H_c + H_1 + H_{ext}. \tag{45}$$

$H_0$ is one-particle Dirac Hamiltonian plus electromagnetic term:

$$H_0 = H_{em} + \int dr \psi^+(x) h(x) \psi(x), \tag{46}$$

where Dirac Hamiltonian is

$$h(x) = c\boldsymbol{\alpha} \cdot \boldsymbol{p} + \beta mc^2. \tag{47}$$

Radiation gauge is used for electromagnetic field. The interaction between electrons is

$$H_c = \frac{e^2}{2} \int dr \int dr' \psi^+(x) \psi(x) \frac{1}{|r-r'|} \psi^+(x') \psi(x'). \tag{48}$$

The interaction between matter and transverse radiation field is

$$H_1 = \frac{e}{c} \int dr \, \boldsymbol{j}(x) \cdot \boldsymbol{A}(x). \tag{49}$$

The definition of four-current is

$$j^\mu(x) = c\psi^+(x) \gamma^0 \gamma^\mu \psi(x), \tag{50}$$

which satisfies continuity equation:

$$\partial_\mu j^\mu(x) = 0. \tag{51}$$

The last term in (45) is the interaction between the current and classical electromagnetic field $A^\mu_{ext}(x)$.

$$H_{ext} = -\frac{e}{c} \int dr \, j_\mu(x) A^\mu_{ext}(x). \tag{52}$$

In these equations, operator order should be normal product.

These equations are relativistic, so that they are valid for both PKE and NKE systems. Nevertheless, when narrating Hohenberg-Kohn theorem, we have to distinguish the PKE and NKE solutions $\psi_{(+)}$ and $\psi_{(-)}$ of relativistic quantum mechanics equation.

If the wave function in Eqs. (44)-(52) is the PKE $\psi_{(+)}$ (NKE $\psi_{(-)}$), then the narration of Hohenberg-Kohn theorem is: Ground state energy is the unique functional of charge and spin densities, and when and only when both densities reach their correct values, the ground state energy reaches its minimum (maximum).

Now, we consider the ground state energy of Hamiltonian (45) in the case of NKE. In this case, the energy reaches maximum.

The current density is the average of $j_\mu(x)$ in the ground state $|G_{(-)}\rangle$:

$$J_{(-)\mu}(x) = \langle G_{(-)} | j_\mu(x) | G_{(-)} \rangle. \tag{53}$$

The four components are not independent of each other, because there is a constraint continuity equation,

$$\partial_\mu J_{(-)}^\mu(x) = 0. \tag{54}$$

The following proof procedure is the same as that for PKE systems. One merely needs to replace the less than sign and minimum there by greater than sign and maximum.

Assume that under the same four-current $J_{(-)\mu}(x)$, there exists another external field $A_{\text{ext}}'^\mu$ such that the Hamiltonian becomes $H'$ and there is correspondingly another ground state $|G'_{(-)}\rangle$ and energy $E'_{(-)} = \langle G'_{(-)} | H' | G'_{(-)} \rangle$.

$$\begin{aligned}
E'_{(-)} &= \langle G'_{(-)} | H' | G'_{(-)} \rangle > \langle G_{(-)} | H' | G_{(-)} \rangle \\
&= \langle G_{(-)} | H - e \int d\mathbf{r}\, j_\mu(x)[A_{\text{ext}}'^\mu(x) - A_{\text{ext}}^\mu(x)] | G_{(-)} \rangle.
\end{aligned} \tag{55}$$

The external field is not operators but numbers, and therefore, it follows that

$$\langle G_{(-)} | \int d\mathbf{r}\, j_\mu(x) A_{\text{ext}}^\mu(x) | G_{(-)} \rangle = \int d\mathbf{r}\, J_{(-)\mu}(x) A_{\text{ext}}^\mu(x). \tag{56}$$

Thus,

$$E'_{(-)} > E_{(-)} - e \int d\mathbf{r}\, J_{(-)\mu}(x)[A_{\text{ext}}'^\mu(x) - A_{\text{ext}}^\mu(x)]. \tag{57}$$

Note that by assumption, in the $|G'_{(-)}\rangle$, $J_{(-)\mu}(x) = \langle G'_{(-)} | j_\mu(x) | G'_{(-)} \rangle$ remains unchanged. Exchanging the quantities with and without prime, we obtain

$$E_{(-)} > E'_{(-)} - e \int d\mathbf{r}\, J_{(-)\mu}(x)[A_{\text{ext}}^\mu(x) - A_{\text{ext}}'^\mu(x)]. \tag{58}$$

It is seen that we meet again that $E + E' > E + E'$, which is absurdity.

This demonstrates that the external field $A_{\text{ext}}^\mu$ must be the unique functional of current density $J_{(-)\mu}$. The Hamiltonian $H$ is in turn determined by $A_{\text{ext}}^\mu$. The conclusion is that the ground state energy is the unique functional of $J_{(-)\mu}$. We can put

down

$$E = E[J]. \tag{59}$$

Integration of continuity equation in three-dimensional space results in

$$\int d\mathbf{r} J^0_{(-)}(x) = \langle G_{(-)} | \int d\mathbf{r}\, j(x) | G_{(-)} \rangle = \text{const.}. \tag{60}$$

This equation corresponds to the normalization condition (35) that the total particle number is $N$.

Next, we prove that when $J_{(-)\mu}(x)$ reaches correct values, $E_{(-)} = E_{(-)}[J_{(-)}]$ reaches its maximum. We define

$$F[J] = \langle G | H_0 + H_c + H_1 | G \rangle. \tag{61}$$

Hence,

$$E_{(-)}[J_{(-)}] = F_{(-)}[J_{(-)}] - e \int d\mathbf{r}\, J_{(-)\mu}(x) A^\mu_{\text{ext}}(x). \tag{62}$$

$E_{(-)}[J_{(-)}]$ is the unique functional of $J_{(-)}$, and either is $F_{(-)}[J_{(-)}]$.

If $J_{(-)}$ changes to be $J'_{(-)}$ and the ground state changes to be $G'$, but $A^\mu_{\text{ext}}$ remains unchanged, then

$$E_{(-)}[G'_{(-)}] = F_{(-)}[J'_{(-)}] - e \int d\mathbf{r}\, J'_{(-)\mu}(x) A^\mu_{\text{ext}}(x). \tag{63}$$

Since now $|G'_{(-)}\rangle$ is not a correct ground state, it should be

$$E_{(-)}[G'_{(-)}] < E_{(-)}[J_{(-)}] = F_{(-)}[J_{(-)}] - e \int d\mathbf{r}\, J_{(-)\mu}(x) A^\mu_{\text{ext}}(x). \tag{64}$$

Therefore, when current density reaches correct value, energy functional reaches its maximum, the condition being that the continuity equation is satisfied, i.e., total particle number is conserved.

One may write down concretely the four-current on the right hand side of Eq. (64). They are listed below.[25]

$$j^0_{(-)}(x) = c\psi^+_{(-)}(x)\psi(x)_{(-)}. \tag{65a}$$

$$j^k_{(-)}(x) = j^k_{(-)A}(x) + \frac{1}{2m}\varepsilon^{kjl}\frac{\partial}{\partial x^j}(\psi^+_{(-)}\sigma^l\psi_{(-)}) - \frac{i}{2mc}\frac{\partial}{\partial t}(\psi^+_{(-)}\alpha^k\psi_{(-)}). \tag{65b}$$

In these equations, the following functional are defined. In Eq. (65a), particle density is defined by

$$\rho_{(-)}(x) = \langle G | \psi^+_{(-)}(x)\psi_{(-)}(x) | G \rangle. \tag{66a}$$

The first term in Eq. (65b) is regular current density caused by electromagnetic potential. For the second term of (65b), one defines spin density vector

$$s^l_{(-)}(x) = \langle G | \psi^+_{(-)}\sigma^l\psi_{(-)} | G \rangle. \tag{66b}$$

The functional corresponding to the last term in (65b) is

$$g_{(-)}^l(x) = -\frac{i}{2m}\langle G|\psi_{(-)}^+ \alpha^l \psi_{(-)}|G\rangle. \tag{66c}$$

The contribution from the spin density part can be written as

$$\frac{e}{2m}\int d\mathbf{r}\,\varepsilon^{kjl}\frac{\partial s_{(-)}(x)}{\partial x^j}A_{ext,k}(x) = -\frac{e}{2m}\int d\mathbf{r}\,\mathbf{s}_{(-)}(x)\cdot\mathbf{B}_{ext}(x), \tag{67}$$

where $\mathbf{s}(x)$ is ordinary spin density and $\mathbf{B}_{ext} = \nabla\times\mathbf{A}_{ext}$ is magnetic induction density. After substitution of Eqs. (65)-(67) into (62), the energy functional becomes

$$E[J] = F[J] + \int d\mathbf{r}[\rho_{(-)}(x)V_{ext}(x) + \frac{e}{2m}\mathbf{s}(x)\cdot\mathbf{B}_{ext}(x) \\ -e\mathbf{J}_{(-)A}(x)\cdot\mathbf{A}(x) + \frac{e}{2m}\frac{d\mathbf{g}_{(-)}(x)}{dt}\cdot\mathbf{A}_{ext}(x)], \tag{68}$$

where $V_{ext}(x) = -e\phi(x)$ is a scalar potential, which is also time component of four-potential $A$. Equation (68) is the relativistic generalization of (42).

### C. Finite negative temperature theory

Here, we consider the case of low momentum.

For PKE systems, Hohenberg-Kohn theorem has been generalized to the case of finite temperature.[26] The idea was that at a fixed temperature, the grand potential of the system is a unique functional of density, and that when the density reaches correct value, grand potential reaches its minimum.

Now let us discuss NKE systems in the way almost the same as PKE systems. The following discuss imitates [27].

The NT and volume are fixed, but particle number is variable. In equilibrium, the chemical potential is fixed and grand potential $\Omega$ reaches maximum. The expression of grand potential is

$$\Omega_{(-)} = -\frac{1}{\beta_{(-)}}\ln\text{tr}[e^{-\beta_{(-)}(H_{(-)}-\mu_{(-)}N)}]. \tag{69}$$

Density matrix operator is

$$\rho_{(-)} = \frac{e^{-\beta_{(-)}(H_{(-)}-\mu_{(-)}N)}}{\text{tr}[e^{-\beta_{(-)}(H_{(-)}-\mu N_{(-)})}]}. \tag{70a}$$

It is normalized,

$$\text{tr}\rho_{(-)} = 1. \tag{70b}$$

The ensemble average of a physical quantity $A$ is that

$$\langle A\rangle = \text{tr}(A\rho_{(-)}). \tag{71}$$

The particle density can be evaluated by

$$\rho_{(-)}(r) = \mathrm{tr}[\rho \psi_{(-)}^+(r,t)\psi_{(-)}(r,t)]. \tag{72}$$

Here we use $\rho_{(-)}(r)$ to represent particle density, which is a number, and use $\rho$ to represent density matrix, which is an operator.

Suppose that there is a following functional of the density matrix.

$$\omega_{(-)}[\rho_T] = \mathrm{tr}\{\rho_T (H_{(-)} - \mu_{(-)} N + \beta_{(-)}^{-1} \ln \rho_T)\}, \tag{73}$$

where Hamiltonian is

$$H_{(-)} = K_{(-)} + U + V. \tag{74}$$

By Eq. (36),

$$V - \mu_{(-)} N = \int d\boldsymbol{r}(v(\boldsymbol{r}) - \mu_{(-)})\psi_{(-)}^+(\boldsymbol{r})\psi_{(-)}(\boldsymbol{r}). \tag{75}$$

When the density matrix $\rho_T$ in (73) becomes correct density matrix $\rho$ in the system, from (70),

$$\begin{aligned}
\omega_{(-)}[\rho] &= \mathrm{tr}\{\rho(H_{(-)} - \mu_{(-)} N + \beta_{(-)}^{-1} \ln \rho)\} \\
&= \mathrm{tr}\{\rho(-\beta^{-1} \ln \mathrm{tr}[e^{-\beta_{(-)}(H_{(-)} - \mu_{(-)} N)}])\} \\
&= -\beta_{(-)}^{-1} \mathrm{tr}(\rho) \ln \mathrm{tr}[e^{-\beta_{(-)}(H_{(-)} - \mu_{(-)} N)}] \\
&= -\beta_{(-)}^{-1} \ln \mathrm{tr}[e^{-\beta_{(-)}(H_{(-)} - \mu_{(-)} N)}] = \Omega_{(-)}[\rho].
\end{aligned} \tag{76}$$

The last step is just the definition (69).

The formulas above are formally the same as those of PKE systems. Two points should be noted. One is that temperature of a PKE system is positive, while that of a NKE system is negative. The other is that the grand potential of a PKE system is negative and reaches its minimum at equilibrium, while that of a NKE system is positive and reaches its maximum at equilibrium.

Consequently, one needs just exchanging the greater than and less than signs in the proof process for a PKE system.

Since in equilibrium, grand potential reaches its maximum, when the density is not the value of equilibrium, the corresponding grand potential should be less than equilibrium grand potential:

$$\Omega_{(-)}[\rho_T] < \Omega_{(-)}[\rho]. \tag{77}$$

Now, assume that external field and chemical potential are shifted, and their difference becomes $v'(r) - \mu'_{(-)}$, but density $\rho_{(-)}(r)$ remains correct. Nevertheless, because $v'(r) - \mu'_{(-)}$ varies, the density matrix as an operator changes, $\rho' \neq \rho$. Equation (73) becomes

$$\omega'_{(-)}[\rho'_{(-)}] = \text{tr}\{\rho'(H'_{(-)} - \mu'_{(-)}N + \beta^{-1}_{(-)}\ln\rho')\}$$
$$< \int d\mathbf{r}[v'(\mathbf{r}) - \mu'_{(-)} - (v(\mathbf{r}) - \mu_{(-)})]\rho_{(-)}(\mathbf{r}) + \omega_{(-)}[\rho_{(-)}]. \qquad (78)$$

Exchanging the functional and operators with and without prime in (78) results in

$$\omega_{(-)}[\rho_{(-)}] < \int d\mathbf{r}[v(\mathbf{r}) - \mu_{(-)} - (v'(\mathbf{r}) - \mu'_{(-)})]\rho'_{(-)}(\mathbf{r}) + \omega'_{(-)}[\rho'_{(-)}]. \qquad (79)$$

From these two equations we have

$$\omega[\rho_{(-)}] + \omega'[\rho'_{(-)}] < \omega[\rho_{(-)}] + \omega'[\rho'_{(-)}]. \qquad (80)$$

This is absurdity. Obviously the assumption of $\rho' \neq \rho$ was not correct. Since the density matrix $\rho$ corresponding to equilibrium $\rho_{(-)}(\mathbf{r})$ is unique, and $\rho_{(-)}(\mathbf{r})$ in principle determines $v(\mathbf{r}) - \mu_{(-)}$ and $H_{(-)} - \mu_{(-)}N$, we can say that density matrix $\rho$ is a functional of $\rho_{(-)}(\mathbf{r})$ which is in fact the diagonal elements of density matrix.

Let us define the following functional.

$$F[\rho] = \text{tr}\{\rho(K + U + \beta^{-1}\ln\rho)\} = K[\rho] + U[\rho] - TS[\rho]. \qquad (81)$$

The three functionals on the right hand side are respectively kinetic energy, interaction energy, and entropy. Equation (73) is rewritten as

$$\omega[\rho] = \int d\mathbf{r}[v(\mathbf{r}) - \mu]\rho(\mathbf{r}) + F[\rho]. \qquad (82)$$

This equation corresponds to expression of grand potential,

$$\Omega = E - TS - \mu N. \qquad (83)$$

Finally, it follows from (77) that

$$\omega[\rho_T] \leq \omega[\rho]. \qquad (84)$$

Therefore, at the correct density matrix, the grand potential reaches its maximum.

For PKE systems, the generalization of Hohenberg-Kohn theorem to the case of finite temperature and relativity has been done.[28] The case of NKE systems can imitate the discussion, but we do not do the work here.

## IV. KOHN-SHAM EQUATIONS

### A. Zero temperature theory

For a PKE electronic system, Kohn and Sham[29] derived a set of self-consistent equations that single electrons satisfied. For NKE electron systems, the same thing can be done.

The evaluation of the ground state energy of the system starts from Eq. (42). The free energy functional in (42) can be more explicitly separated into three terms: NKE, Coulomb repulsive energy between electrons, and exchange-correlation energy.

$$F[\rho(r)] = K[\rho(r)] + \frac{e^2}{2}\int dr \int dr' \frac{\rho(r)\rho(r')}{|r-r'|} + E_{xc}[\rho(r)], \tag{85}$$

where the concrete form of exchange-correlation energy has not been known yet. Particle number conservation provides a condition

$$\int dr \delta\rho(r) = 0. \tag{86}$$

The variation with respect to ground state energy is

$$\int dr \delta\rho(r) \{\frac{\delta k[\rho]}{\delta\rho} + v(r) + e^2 \int dr' \frac{\rho(r')}{|r-r'|} + \frac{\delta\varepsilon_{xc}[\rho]}{\delta\rho}\} = 0, \tag{87}$$

where kinetic energy density and exchange-correlation energy density are defined. With the help of constraint condition (86), we have

$$\frac{\delta k[\rho]}{\delta\rho} + v(r) + e^2 \int dr' \frac{\rho(r')}{|r-r'|} + \frac{\delta\varepsilon_{xc}[\rho]}{\delta\rho} = \mu, \tag{88}$$

where the Lagrange multiplier $\mu$ is just chemical potential. The three potentials can be combined into one which is regarded as an effective potential the electrons feel.

$$v_{eff}(r) = v(r) + e^2 \int dr' \frac{\rho(r')}{|r-r'|} + v_{xc}(r), \tag{89}$$

where "exchange-correlation potential" is defined:

$$v_{xc}(r) = \frac{\delta\varepsilon_{xc}[\rho]}{\delta\rho}. \tag{90}$$

The exchange-correlation potential may be nonlocal. The concrete form of kinetic density is determined in the following way. Suppose that there are $N$ particles in the system. We choose a set of single particle wave functions, and they are employed to express the real density:

$$\sum_{i=1}^{N} |u_i(r)|^2 = \rho(r). \tag{91}$$

Equation (91) is a characteristic of this theory. For PKE systems, this is exactly provable.

Note that in general, the orbitals $u_i^*(r)$ are complex numbers. With the form of (91), the NKE functional can be written as

$$K[\rho(r)] = -\sum_{i=1}^{N} \int dr \nabla u_i^*(r) \cdot \nabla u_i(r) = \sum_{i=1}^{N} \int dr u_i^*(r) \nabla^2 u_i(r). \tag{92}$$

Equation (92) is not rigorously, but merely approximately, equal to real NKE density. Their difference is merged into the exchange-correlation energy. Then, we take variation of the ground state energy with respect to $u_i^*(r)$, so as to get equations that

single particles satisfy:

$$[\nabla^2 + v_{\text{eff}}(\mathbf{r})]u_i(\mathbf{r}) = \varepsilon_i u_i(\mathbf{r}). \tag{93}$$

In this equation the single particle operator in square brackets is Hermitian, so that the solved orbitals are orthogonal to each other.

We manipulate (93) by multiplying by $u_i^*(\mathbf{r})$ and integrating in the whole space. Assuming that each orbital is normalized, we then take summation with respect to $i$.

$$\int d\mathbf{r}[\sum_{i=1}^{N} u_i^*(\mathbf{r})\nabla^2 u_i(\mathbf{r}) + v_{\text{eff}}(\mathbf{r})\rho(\mathbf{r})] = \sum_{i=1}^{N} \varepsilon_i, \tag{94}$$

where Eq. (91) is employed. Substituting (85) into (42), we get the expression of ground state energy:

$$E_G[\rho] = \int d\mathbf{r} v(\mathbf{r})\rho(\mathbf{r}) + K[\rho(\mathbf{r})] + \frac{e^2}{2}\int d\mathbf{r}\int d\mathbf{r}' \frac{\rho(\mathbf{r})\rho(\mathbf{r}')}{|\mathbf{r}-\mathbf{r}'|} + E_{\text{xc}}[\rho(\mathbf{r})]. \tag{95}$$

Comparison of (94) and (95) leads to

$$E_G[\rho] = \sum_{i=1}^{N} \varepsilon_i - \frac{e^2}{2}\int d\mathbf{r}\int d\mathbf{r}' \frac{\rho(\mathbf{r})\rho(\mathbf{r}')}{|\mathbf{r}-\mathbf{r}'|} + E_{\text{xc}}[\rho(\mathbf{r})] - \int d\mathbf{r} v_{\text{xc}}(\mathbf{r})\rho(\mathbf{r}). \tag{96}$$

It is seen that the ground state energy is the sum of the highest energy of all single particle orbits plus the difference in the exchange correlation energy, and the Coulomb energy between electrons should be subtracted. In the course of the summing the single-particle energy, the Coulomb energy has been counted twice. Equation (96) is formally the same as that of a PKE system.

For PKE systems, Lam and Platzman[30] proved that momentum density function $N_p$, i.e., the number of electrons with momentum $\mathbf{p}$, is

$$N_p = \sum_{i=1}^{N} |\langle \mathbf{p} | u_i \rangle|^2 + \text{correction}, \tag{97}$$

where the correction term concerns the difference of the distributions of interactive and non-interactive electrons. For NKE systems, it is believed that this equation is also valid.

Now we consider spin. After quantization axis is selected, electric density is divided into two parts: spin up and spin down.

$$\rho = \rho_\sigma + \rho_{-\sigma}. \tag{98}$$

Exchange-correlation energy depends on both parts:

$$E_{\text{ec}} = E_{\text{xc}}[\rho_\sigma, \rho_{-\sigma}]. \tag{99}$$

Equation (91) should be changed to be

$$\sum_{i=1}^{N_\sigma} |u_{i\sigma}(\mathbf{r})|^2 = \rho_\sigma(\mathbf{r}). \tag{100}$$

The formulism (85)-(96) still applies, except that in exchange-correlation energy, the variation is taken with respect to the density of spin components. Thus, Eq. (90) is rewritten as

$$v_{xc\sigma}(\mathbf{r}) = \frac{\delta \varepsilon_{xc}[\rho_\sigma, \rho_{-\sigma}]}{\delta \rho_\sigma}. \tag{101}$$

Correspondingly, Eq. (93) that single particles satisfy is recast to be

$$[\nabla^2 + v(\mathbf{r}) + e^2 \int d\mathbf{r}' \frac{\rho(\mathbf{r}')}{|\mathbf{r}-\mathbf{r}'|} + v_{xc\sigma}(\mathbf{r})] u_{i\sigma}(\mathbf{r}) = \varepsilon_{i\sigma} u_{i\sigma}(\mathbf{r}). \tag{102}$$

The ground state energy is changed from (96) to be

$$E_G[\rho] = \sum_{i=1}^N \varepsilon_i - \frac{e^2}{2}\int d\mathbf{r}\int d\mathbf{r}' \frac{\rho(\mathbf{r})\rho(\mathbf{r}')}{|\mathbf{r}-\mathbf{r}'|} + E_{xc} - \sum_\sigma \int d\mathbf{r} v_{xc}(\mathbf{r})\rho_\sigma(\mathbf{r}), \tag{103}$$

which has the same form as that of a PKE system.

**B. Finite negative temperature theory**

Now, we generalize the formulism to the case of finite NT. We still use the three functionals defined by Eqs. (81)-(83) which are copied here:

$$F[\rho] = \mathrm{tr}\{\rho(K+U+\beta^{-1}\ln\rho)\} = K[\rho] + U[\rho] - TS[\rho]. \tag{104}$$

The right hand side is the functionals of kinetic energy, interactive energy and entropy. Equation (82) is written as

$$\omega[\rho] = \int d\mathbf{r}[v(\mathbf{r}) - \mu]\rho(\mathbf{r}) + F[\rho]. \tag{105}$$

The grand potential is

$$\Omega = E - TS - \mu N. \tag{106}$$

It follows that

$$\begin{aligned}\Omega[\rho] &= \int d\mathbf{r}[v(\mathbf{r})-\mu]\rho(\mathbf{r}) + F[\rho] \\ &= \int d\mathbf{r}[v(\mathbf{r})-\mu]\rho(\mathbf{r}) + T[\rho] + U[\rho] - TS[\rho] \\ &= K[\rho] - TS[\rho] + \int d\mathbf{r}[v(\mathbf{r})-\mu]\rho(\mathbf{r}) + \frac{e^2}{2}\int d\mathbf{r}\int d\mathbf{r}' \frac{\rho(\mathbf{r})\rho(\mathbf{r}')}{|\mathbf{r}-\mathbf{r}'|} + \Omega_{xc}.\end{aligned} \tag{107}$$

In the end, exchange-correlation energy is added. The first two terms, kinetic energy and entropy, constitute Gibbs free energy, i.e.,

$$G_S[\rho] = K[\rho] - TS[\rho]. \tag{108}$$

The variation of the grand potential (108) is taken with respect to the density. Because chemical potential has been included, Eq. (86) is needless.

$$\frac{\delta G_S[\rho]}{\delta \rho(\mathbf{r})} + v(\mathbf{r}) + e^2 \int d\mathbf{r}' \frac{\rho(\mathbf{r}')}{|\mathbf{r}-\mathbf{r}'|} + v'_{xc}[\rho] - \mu = 0, \tag{109a}$$

where

$$v'_{xc}[\rho] = \frac{\delta\Omega_{xc}[\rho]}{\delta\rho(r)}. \tag{109b}$$

Now, exchange-correlation energy depends on NT. Equation (109) can be compared to (88) and (90). The Gibbs free energy density in the former replaces the kinetic energy density in the latter.

Again like Eq. (89), the three terms in Eq. (109) can be combine to be an effective potential:

$$v_{eff} = v(r) + e^2 \int dr' \frac{\rho(r')}{|r-r'|} + v'_{xc}[\rho]. \tag{110}$$

In this way, Eq. (109) is formally simplified to be

$$\frac{\delta G_S[\rho]}{\delta\rho(r)} + v_{eff}(r) - \mu = 0. \tag{111}$$

For a non-interactive system, the grand potential is[4]

$$\Omega = -k_B T \sum_i \ln(1 + e^{-\beta(\varepsilon_i - \mu)}). \tag{112}$$

The particles in the $i$th state observes Fermi-Dirac distribution:

$$n_i = \frac{1}{e^{\beta(\varepsilon_i - \mu)} + 1}. \tag{113}$$

Entropy can be expressed by

$$S = -(\frac{\partial\Omega}{\partial T})_{\mu,V} = -k_B \sum_i [n_i \ln n_i + (1 - n_i)\ln(1 - n_i)]. \tag{114}$$

These two equations look as if the copies of those of PKE systems.[4]

In finite NT, infinite orbitals ought to be calculated. The condition that particle number conserves is

$$\sum_{i=1}^{\infty} |u_i(r)|^2 n_i = \rho(r). \tag{115}$$

That is to say, in each orbital of (91), a Fermi-Dirac distribution function must be multiplied.

When particle number is a constant, we merely need to consider Helmholtz free energy $A[\rho]$.

$$A[\rho] = \Omega[\rho] + \mu N, \tag{116}$$

see Eqs. (106) and (107). We move the term $\mu N$ to the left hand side, which is equivalent to subtrate $\mu N$ on both sides. Then, we take variation with respect to orbitals. Please note that although entropy depends on density, it does not explicitly

contain orbital. Therefore, the variation results in

$$[\nabla^2 + v(\mathbf{r}) + e^2 \int d\mathbf{r}' \frac{\rho(\mathbf{r}')}{|\mathbf{r}-\mathbf{r}'|} + v_{xc\sigma}(\mathbf{r})]u_i(\mathbf{r}) = \varepsilon_i u_i(\mathbf{r}). \quad (117)$$

This formula is basically the same as (93). Now, Eqs. (113), (115) and (117) are solve simultaneously by self-consistent calculation.

Let us investigate the following matrix element:

$$\rho_1(\mathbf{r}',\mathbf{r}) = N \int d\mathbf{r}_2 d\mathbf{r}_3 \cdots d\mathbf{r}_N \Psi(\mathbf{r}',\mathbf{r}_2,\cdots,\mathbf{r}_N) \Psi^*(\mathbf{r},\mathbf{r}_2,\cdots,\mathbf{r}_N). \quad (118)$$

For this element, Lowdin[31] introduced the following expression,

$$\rho_1(\mathbf{r}',\mathbf{r}) = \sum_{i=1}^{\infty} \phi_i(\mathbf{r}')\phi_i^*(\mathbf{r}) n_i, \quad (119)$$

where $\phi_i$ was called natural orbital. In terms of the natural orbital, kinetic energy functional is of a simple form,

$$K[\rho(\mathbf{r})] = -\int d\mathbf{r}[\nabla_{\mathbf{r}'} \cdot \nabla_{\mathbf{r}} \rho(\mathbf{r}',\mathbf{r})]_{\mathbf{r}'=\mathbf{r}} = \sum_{i=1}^{\infty} n_i \int d\mathbf{r} \phi_i^*(\mathbf{r}) \nabla^2 \phi_i(\mathbf{r}). \quad (120)$$

Let us discuss the meaning of single particle eigenvalue $\varepsilon_i$. After the orbitals are solved from Eq. (117) and substituted into total energy functional, the total energy $E_T$ can be evaluated. It follows that

$$\varepsilon_i = \partial E_T / \partial n_i. \quad (121)$$

Equation (121) means that the occupation number $n_i$ can vary continuously. This equation applies to an infinitely large system, and is valid in statistical sense. In order to prove Eq. (121), the grand potential at finite NT should be taken into account. The grand potential (107) is recast into

$$\Omega[\rho] = K[\rho] - TS + \int d\mathbf{r}[v(\mathbf{r}) - \mu]\rho(\mathbf{r}) + \frac{e^2}{2} \int d\mathbf{r} \int d\mathbf{r}' \frac{\rho(\mathbf{r})\rho(\mathbf{r}')}{|\mathbf{r}-\mathbf{r}'|} + \Omega_{xc}. \quad (122)$$

Here, entropy has been chosen as rigorous. If the rigorous one differs from that in Eq. (107), then the difference is merged into exchange-correlation functional. Now the internal energy is put down as

$$E_T = \Omega + TS + \mu N$$
$$= \sum_i n_i \int d\mathbf{r} u_i^*[\nabla_i^2 + v(\mathbf{r})]u_i(\mathbf{r}) + \frac{e^2}{2} \sum_{i,j}^{\infty} n_i n_j \int d\mathbf{r} \int d\mathbf{r}' \frac{|u_i(\mathbf{r})|^2 |u_j(\mathbf{r}')|^2}{|\mathbf{r}-\mathbf{r}'|} + \Omega_{xc}. \quad (123)$$

We take derivative with respect to $n_i$. The derivative of exchange-correlation energy is

$$\frac{\partial \Omega_{xc}}{\partial n_i} = \int d\boldsymbol{r} \frac{\partial \Omega_{xc}[\rho]}{\partial \rho(\boldsymbol{r})} \frac{\partial \rho(\boldsymbol{r})}{\partial n_i} = \int d\boldsymbol{r} \, |u_i(\boldsymbol{r})|^2 \, v'_{xc}(\boldsymbol{r}). \tag{124}$$

Then,

$$\begin{aligned}\frac{\partial E_T}{\partial n_i} &= \int d\boldsymbol{r} u_i^* [\nabla_i^2 + v(\boldsymbol{r}) + e^2 \int d\boldsymbol{r}' \frac{\rho(\boldsymbol{r}')|u_j(\boldsymbol{r}')|^2}{|\boldsymbol{r} - \boldsymbol{r}'|} + v'_{xc}(\boldsymbol{r})] u_i(\boldsymbol{r}) \\ &= \int d\boldsymbol{r} u_i^* \varepsilon_i u_i(\boldsymbol{r}) = \varepsilon_i,\end{aligned} \tag{125}$$

where Eqs. (115) and (117) have been employed. Thus, Eq. (121) is proven.

Subsequently, the change of total energy is

$$dE = \sum_i \frac{\partial E}{\partial n_i} dn_i = \sum_i \varepsilon_i dn_i. \tag{126}$$

Based on the above formulism, the variation of the grand potential is

$$d\Omega = dE - TdS - \mu dN = \sum_i (\varepsilon_i - \mu + k_B T \ln \frac{n_i}{1-n_i}) dn_i = 0. \tag{127}$$

At equilibrium, the grand potential reaches its maximum, so that the variation should be zero. It is seen that

$$n_i = \frac{1}{e^{\beta(\varepsilon_i - \mu)} + 1}, \tag{128}$$

which is just Fermi-Dirac distribution. This also confirms that electrons indeed occupy energy states that are solved from Eq. (117).

At zero temperature limit, exchange-correlation functional degrades to the value in the ground state,

$$\lim_{T \to 0} \Omega_{xc}[\rho] = E_{xc}[\rho]. \tag{129}$$

The exchange-correlation energy of the ground state is believed exact. Equation (125) is also formally exact. At zero temperature, since the distribution (128) degrades to the step function, Eq. (115) degrades to (91), and the integration of both are total particle number, i.e., the constraint condition at zero temperature. Inside and outside the Fermi sphere, the occupation numbers are respectively 1 and 0, and at the Fermi energy level the occupation number is a fraction.

## V. HARTREE-FOCK SELF-CONSISTENT EQUATIONS

For PKE systems, there have been famous Hartree-Fock self-consistent equations.[32–34] For NKE system the same equations can be established.

The wave function of Fermions is written in the form of Slater determinant,

$$\Psi(1,2,\cdots,N) = \frac{1}{\sqrt{N!}} \begin{vmatrix} \varphi_1(q_1) & \varphi_1(q_2) & \cdots & \varphi_1(q_N) \\ \varphi_2(q_1) & \varphi_2(q_2) & \cdots & \varphi_2(q_N) \\ \vdots & \vdots & \ddots & \vdots \\ \varphi_N(q_1) & \varphi_N(q_2) & \cdots & \varphi_N(q_N) \end{vmatrix}. \tag{130}$$

Hamiltonian includes two parts: single-particle Hamiltonian and two-body interaction Hamiltonian.

$$H = \sum_i h_i + \frac{1}{2}\sum_{i\neq j} V_{ij}. \tag{131}$$

Both parts are Hermitian. The average of Hamiltonian $\bar{H} = (\Psi, H\Psi)$ in the wave function (130) is

$$\bar{H} = \sum_i \int dx_i \varphi_i^*(r) h_i \varphi_i(r) + \frac{1}{2}\sum_{i\neq j} \iint dx_i dx_j \varphi_i^*(r)\varphi_j^*(r')V_{ij}\varphi_i(r)\varphi_j(r') \\ - \frac{1}{2}\sum_{i\neq j} \iint dx_i dx_j \varphi_i^*(r)\varphi_j^*(r')V_{ij}\varphi_i(r')\varphi_j(r). \tag{132}$$

We take variation of this average.

$$\delta\bar{H} = \sum_i \int dr[\delta\varphi_i^*(r)h_i\varphi_i(r) + \varphi_i^*(r)h_i\delta\varphi_i(r)] \\ + \sum_{i\neq j} \iint drdr'[\delta\varphi_i^*(r)\varphi_j^*(r')V_{ij}\varphi_i(r)\varphi_j(r') + \varphi_i^*(r)\varphi_j^*(r')V_{ij}\delta\varphi_i(r)\varphi_j(r')] \\ - \sum_{i\neq j} \iint drdr'[\delta\varphi_i^*(r)\varphi_j^*(r')V_{ij}\varphi_i(r')\varphi_j(r) + \varphi_i^*(r)\varphi_j^*(r')V_{ij}\delta\varphi_i(r')\varphi_j(r)]. \tag{133}$$

Assume that single particle wave functions are normalized. The constraint conditions are

$$(\varphi_i, \varphi_i) = 1, i = 1, 2, \cdots; \delta\bar{H} - \sum_i \varepsilon_i \delta(\varphi_i, \varphi_i) = 0. \tag{134}$$

Here, the $(a,b)$ means the inner product of the $a$ and $b$. Under these conditions, the self-consistent equations are obtained.

$$h_i\varphi_i(r) + \sum_{j(\neq i)}\int dr'[\varphi_j^*(r')V_{ij}\varphi_i(r)\varphi_j(r') - \sum_{j(\neq i)}\int dr'\varphi_j^*(r')V_{ij}\varphi_i(r')\varphi_j(r) \\ = \varepsilon_i\varphi_i(r). \tag{135}$$

The Lagrange multipliers are just the eigen energies of solved orbitals. The last term on the left hand side is exchange energy. Without this term, the equations degrade to Hartree equations. For Hartree-Fock self-consistent equations, there was a Koopmann's theorem[35] which said that the eigenvalue $\varepsilon_i$ corresponding to single particle orbital $\varphi_i$ was the minus of the energy that was needed to remove an electron from orbital $\varphi_i$. The difference of two eigenvalues $\varepsilon_i - \varepsilon_j$ was the energy needed when an electron transited from orbital *i* to *j*. In considering that, the relaxation generated by electron transition was neglected. This theorem also applies to the NKE case.

The process above is the same as that for PKE systems. The only thing one should keep in mind is that in one-body operators, there is no minus sign in kinetic energy term, and the nuclei and electrons have the same kind of change so that there is repulsive potential between them.

Since the single-particle operator in (135) is Hermitian, the solved orbital are orthogonal to each other, $(\varphi_i, \varphi_j) = \delta_{ij}$.

If the wave function of the system is simply the product of orbitals instead of Slater determinant (130), then exchange effect between particles is ignored. We then drop the terms embodying exchange effect in Eqs. (132), (133) and (135). The formulism is simplified to be those of Hartree self-consistent equations.

It is well known that one pair of electrons with opposite spins can occupy one orbital. This pair of electrons are spatially intimately close to each other. This embodies the effect of strong correlation. This effect is difficult to deal with. It was once suggested that two-particle wave functions might correctly take into account the effect and the Hartree-Fock self-consistent equations for the two-particle functions were established.[36] It is believed that for NKE system, the effect of strong correlation can be deal with in the same way.

There was a more detailed theory based on many-body Green's function theory which lead to the Hartree-Fock self-consistent equations for both zero temperature and finite temperature.[37] Similarly, we are able to write the Hartree-Fock self-consistent equations for finite NT. As a matter of fact, this is done by inserting into the second and third terms of Eq. (135) a quantum statistical distribution factor, such as Eq. (128), cf. (13.2.35) in [37].

Finally, the discrepancy between PKE and NKE systems is emphasized. For PKE particle systems, if particles are charged, the interaction between particles is repulsive. Then, both the kinetic energy and interaction energy are positive. There should be a negative external potential such that the system can reach a stable state. For an electron gas, this is the attraction of atomic nuclei. While for NKE particle systems, Coulomb repulsion energy is positive and kinetic energy is negative. In this case, even without an external potential, the system is possible to be stabilized. That is to say, the external potential in the above formula can be zero. Even if the external trend is positive, it may still be possible for a system to reach a balance.

## VI. SUMMARY

The formulas of the many-body theories of negative kinetic energy systems is introduced. They are Thomas-Fermi method, Hohenberg-Kohn theorem, Khon-Sham self-consistent equations that are based on the Hohenberg-Kohn theorem, and Hartree-Fock self-consistent equations. In each case, both zero-temperature and finite-negative-temperature formulas are discussed. The theories are established imitating those of PKE system, so that is almost parallel to those for PKE systems. The difference between the PKE and NKE systems is that the former obey energy minimum principle while the latter energy maximum principle. Since NKE objects are dark ones, the theories presented in this paper provide the means of studying dark matters in detail.

**Acknowledgments.** This work is supported by the National Natural Science Foundation of China under Grant No. 12234013, and the National Key Research and Development Program of China Nos. 2018YFB0704304 and 2016YFB0700102.

## APPNDIX: EXPERIMENTAL SCENARIO FOR COLLISION BETWEEN ELECTRONS AND TUNNELING ELECTRONS

First of all, we raise a question.

In a scanning tunneling microscopy (STM), the gap between the tip and sample surface is empty. According to QM textbooks, when an electric current flows through the gap, an electron in the gap has an energy $E$ that is less than potential $V$, $E < V$. It is believed that the electron tunnels through the gap. Because $E < V$, the kinetic energy is negative, and, by common sense, the electron's momentum will be an imaginary number. It is possible to let particles, such as photons and electrons, incident to the gap and collide with the tunneling electrons. During a collision process, the total energy and total momentum of the colliding particles are conserved.

Our question is: is there any momentum transfer between the colliding particles?

The incident particle has a PKE and the tunneling electron is believed to have an imaginary momentum. Because the total momentum is conserved, both the real and imaginary parts of the total momentum should be conserved. It is thus concluded that the real momentum of the incident particle remains unchanged during the collision, and the imaginary momentum of the tunneling electron remains as well. The answer to the question is that there is no momentum transfer between the particles. That is to say, to the incident particles, the tunneling current is transparent. This answer must be verified by experiments.

The author's point of view is that the tunneling electron's momentum is not imaginary but still real, although its kinetic energy is negative.[1] Consequently, we can write down the equation of conservation of the total momentum. Before and after collision, the incident particle has PKE and the tunneling electron has NKE. In [1], experimental scenarios were suggested which were to use photons to collide NKE electrons.

In this appendix, we suggest a new scenario that is to use PKE electeons to collide tunneling electrons. This scenario is expected be more easily implemented compared to the previous ones using photons.

In each of Figs. 1 and 2 below, two electrons collide with each other. Their masses are labeled by $m_1$ and $m_2$, respectively, and $m_1 = m_2$. $\boldsymbol{p}$ represents momentum and $v$ velocity. The unprimed $p$ and $v$ are the quantities before the collision, and primed quantities after collision. The subscripts 1 and 2 respectively label the quantities belonging to electrons $m_1$ and $m_2$.

The electron $m_1$ is of PKE, the momentum and velocity of which have the same direction, and $m_2$ of NKE, the momentum and velocity of which have opposite directions.

All the momenta are in one plane. Every collision is elastic, and total momentum and kinetic energy are conserved.

**Experiment 1. A PKE electron impinges a tunneling electron**

The experimental scenario is shown in Fig. 1. From the conservation of momentum and kinetic energy, we have the following equations.

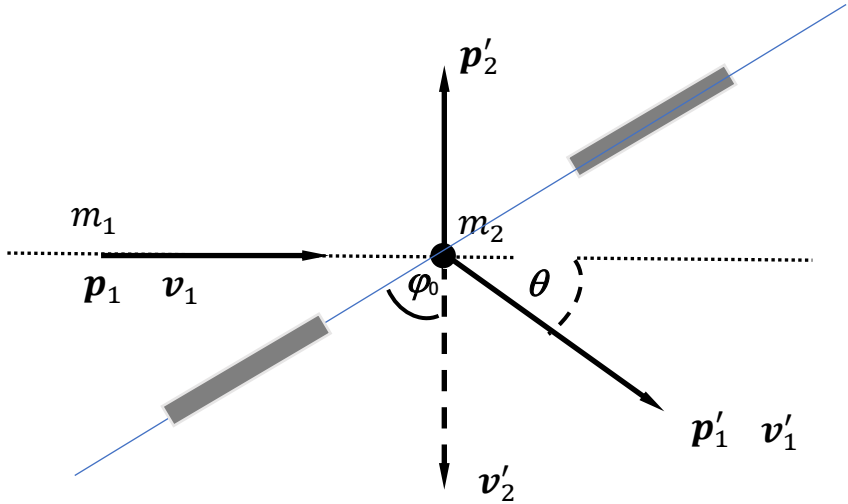

**FIG. 1.** A PKE electron with $p_1$ and $v_1$ is incident from the left. The grey tubes are conductors, which form an angle of $\varphi_0$ with the incident direction. There is a gap between the tubes and electronic current can tunnel through the gap. The incident electron impinges a tunneling electron. It is assumed that after collision, the scattering direction of the $m_1$ is $\theta$ in the lower side and that of $m_2$ is $\varphi$ (not shown) in the upper side of the incident direction. When $p_2 \approx 0$, $p'_2 \perp p_1$, the case of Eq. (A7).

$$p_1 + p_2 \cos\varphi_0 = p'_1 \cos\theta + p'_2 \cos\varphi. \tag{A1}$$

$$p_2 \sin\varphi_0 = p'_1 \sin\theta + p'_2 \sin\varphi. \tag{A2}$$

$$p_1^2 - p_2^2 = p_1'^2 - p_2'^2. \tag{A3}$$

Please note that before and after collision, the tunneling electron is of NKE. There are three equations but four unknown quantities: the outgoing angle and momentum value of the PKE electron $(\theta, p_1')$ and those of the NKE electron $(\varphi, p_2')$. Assuming a fixed $\theta$ angle, we can solve other three quantities as follows.

$$p_1' = \frac{p_1(p_1 + p_2 \cos \varphi_0)}{p_1 \cos \theta + p_2 \cos(\varphi_0 - \theta)}. \tag{A4}$$

$$p_2' = \sqrt{p_2^2 + p_1'^2 - p_1^2}. \tag{A5}$$

$$\tan \varphi = \frac{-p_2 \sin \varphi_0 + p_1' \sin \theta}{p_1 + p_2 \cos \varphi_0 - p_1' \cos \theta}. \tag{A6}$$

If an electron detector is set at the angle $\theta$, the momentum $p_1'$ of the outgoing electron can be detected, the value of which should be in agreement with that calculated by (A4).

The outgoing $m_2$ cannot be detected by any known electron detector, because it is of NKE. A simplest case is interesting: the momentum of the tunneling electron is very low so that it is almost zero. Then, in Eqs. (A1)-(A3) we let $p_2 = 0$ and obtain

$$\varphi = \frac{\pi}{2}, p_1' = \frac{p_1}{\cos \theta}, p_2' = p_1 \tan \theta. \tag{A7}$$

In this case, the outgoing direction of the NKE electron is always perpendicular to the incident direction, $\varphi = \pi/2$, no matter what the $p_2'$ value is. The momentum of the PKE electron after collision is greater than that before collision, $p_1' \geq p_1$. The $m_1$ gains NKE so that the $m_2$ gains PKE. The greater the scattering angle $\theta$, the greater the $p_1'$ value. Figure 1 just depicts the case of $\varphi = \pi/2$.

Although the outgoing NKE electrons cannot be detected directly, their momentum values $p_2'$ in each outgoing direction $\varphi$ can be calculated by Eqs. (A5) and (A6). A special case is Eq. (A7) where there is only one outgoing direction of $m_2$.

The direction of a NKE electron's momentum is opposite to that of its velocity. At each scattering angle $\varphi$, of $p_2'$ we known that the NKE electron moves along the $-p_2'$ direction. This feature can be utilized to conduct experiment 2 below.

**Experiment 2. A NKE electron impinges a PKE electron**

The expetiment scenario is depicted in Fig. 2. An NKE electron is produced by experiment 1 as shown in Fig. 1, and it has velocity $v_2$ and momentum $p_2$. It is incident to impinge an PKE electron. From the conservation of momentum and kinetic energy, we have the following equations.

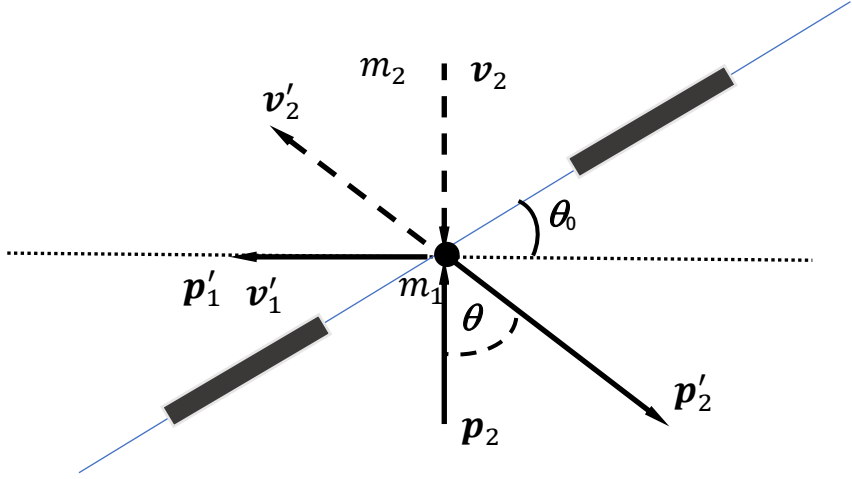

**FIG. 2.** A NKE electron with $p_2$ and $v_2$, produced in experiment 1, is incident from the top. The black tubes are cathode and anode, respectively, which form an angle of $\theta_0$ with the horizontal line. PKE electrons go from the anode to cathode. The incident NKE electron impinges a PKE electron. It is assumed that after collision, the scattering direction of the $m_1$ is $\theta$ in one side and that of $m_2$ is $\varphi$ (not shown) in the other side of the incident direction.

$$p_1 \cos\theta_0 + p_2 = p_1' \cos\theta + p_2' \cos\varphi. \tag{A8}$$

$$p_1 \sin\theta_0 = p_1' \sin\theta - p_2' \sin\varphi. \tag{A9}$$

$$p_1^2 - p_2^2 = p_1'^2 - p_2'^2. \tag{A10}$$

There are three equations but four unknown quantities: the outgoing angle and momentum value of the PKE electron $(\theta, p_1')$ and those of the NKE electron $(\varphi, p_2')$.

For a fixed $\theta$ angle, the momentum value of the outgoing $m_1$ is

$$p_1' = \frac{p_1(p_1 + p_2 \cos\theta_0)}{p_1 \cos(\theta - \theta_0) - p_2 \cos\theta}. \tag{A11}$$

The $(\varphi, p_2')$ of the outgoing $m_2$ can also be calculated but not written here because they are not cared about.

If an electron detector is set at the angle $\theta$, the momentum $p'_1$ of the outgoing electron can be detected, the value of which should be in agreement with that calculated by (A11).

A simplest case is that the momentum of the PKE electrons are so low that it is almost zero. Then, in Eq. (A8)-(A10) we let $p_1 = 0$ and obtain $\theta = \pi/2$. That is to say, there is a relation $p'_1 \perp p_2$. Figure 2 just depicts this case. An electron detector must be placed at the direction $\theta = \pi/2$ such that it can detect outgoing electrons with different momentum values. At any other direction, there is no outgoing PKE electron.

Experiment 1 can be done independently, while experiment 2 cannot. In order to do experiment 2, the devices in Figs. 1 and 2 should be combined. This is because figure 2 requires a NKE electron source, which is from Fig. 1.

**References**
[1]H.-Y. Wang, J. Phys. Commun. **4**, 125004 (2020).
https://doi.org/10.1088/2399-6528/abd00b
[2]E. Schrödinger, Ann. Phy. **386**(18), 109–139 (1926); E. Schrödinger, in *Collected Papers on Wave Mechanics* (Blackie & son Limited: London and Glasow, 1928), pp102–123.
[3]H.-Y. Wang, J. Phys. Commun. **4**, 125010 (2020).
https://doi.org/10.1088/2399-6528/abd340
[4]H.-Y. Wang, J. Phys. Commun. **5**, 055012 (2021).
https://doi.org/10.1088/2399-6528/abfe71
[5]H.-Y. Wang, J. Phys. Commun. **5**, 055018 (2021).
https://doi.org/10.1088/2399-6528/ac016b
[6]H.-Y. Wang, Physics Essays **35**(2), 152 (2022).
http://dx.doi.org/10.4006/0836-1398-35.2.152
[7]H.-Y. Wang, J. of North China Institute of Science and Technology **18**(4), 1 (2021) (in Chinese). http://10.19956/j.cnki.ncist.2021.04.001
[8]H.-Y. Wang, J. of North China Institute of Science and Technology **19**(1), 97 (2022) (in Chinese). http://10.19956/j.cnki.ncist.2022.01.016
[9]H.-Y. Wang, Physics Essays **35**(3), 270 (2022).
http://dx.doi.org/10.4006/0836-1398-35.3.270
[10]H.-Y. Wang, Advanced Studies in Theoretical Physics **16**(3), 131 (2022).
https://doi.org/10.12988/astp.2022.91866
[11]H.-Y. Wang, Physics Essays **36**(2), 140-148 (2023).
http://dx.doi.org/10.4006/0836-1398-36.2.140
[12]H.-Y. Wang, Physics Essays **36**(2), 149-159 (2023).
http://dx.doi.org/10.4006/0836-1398-36.2.149
[13]S. L. Shapiro and S. A. Teukolsky, *Black Holes, White Dwarfs, and Neutron Stars The Physics of Compact Objects* (WILEY-VCH Verlag GmbH & Co. KGaA, Weinheim,

2004).

[14] J. Callaway and N. H. March, Density Functional Methods: Theory and Applications, in *Solid State Physics—advances in Research and Applications, ed. by Henry Ehrenreich, Divid Turnbull and Frederick Seitz, A, Vol. 38* (Academic Press, Inc., Orlando, 1984).

[15] R. E. Marshak and H. A. Bethe, Astrophys. J. **91**, 239 (1940).

[16] T. Sakai, Proc. Phys. Math. Soc. Jpn. **24**, 254 (1942).

[17] R. P. Feynmann, N. Metropolis, and E. Teller, Phys. Rev. **75**, 1561 (1949). https://doi.org/10.1103/PhysRev.75.1561

[18] P. Hohenberg, and W. Kohn, Phys. Rev. **136**, B864 (1964). https://doi.org/10.1103/PhysRev.136.B864

[19] M. Levy, Proc. Natl. Acad. Sci. U. S. A. **76**(12), 6062 (1979). https://www.jstor.org/stable/70178

[20] M. Levy, Phys. Rev. A **26**, 1200 (1982). https://doi.org/10.1103/PhysRevA.26.1200

[21] J. C. Stoddart and N. H. March, Ann. Phys. (New York) **64**(1), 174 (1971). https://doi.org/10.1016/0003-4916(71)90283-1

[22] M. M. Pant and A. K. Rajagopal, Solid State Commun. **10**(12), 1157 (1972). https://doi.org/10.1016/0038-1098(72)90934-9

[23] U. von Barth and L. Hedin, J. Phys. C **5**(13), 1629 (1972). DOI 10.1088/0022-3719/5/13/012

[24] A. K. Rajagopal and J. Callaway, Phys. Rev. B **7**, 1912 (1973). https://doi.org/10.1103/PhysRevB.7.1912

[25] J. Callaway and N. H. March, Density Functional Methods: Theory and Applications, in *Solid State Physics—advances in Research and Applications, ed. by Henry Ehrenreich, Divid Turnbull and Frederick Seitz, A, Vol. 38* (Academic Press, Inc., Orlando, 1984). p151.

[26] N. D. Mermin, Phys. Rev. **137**, A1441 (1965). https://doi.org/10.1103/PhysRev.137.A1441

[27] W. Kohn and P. Vashishta, *Theory of the Ingomogeneous Electron Gas* S. Lundqvist and N. H. March, eds. (Plenum, New York, 1983). Chap. 2.

[28] A. K. Rajagopal, Adv. Chem. Phys. **41**, 59 (1979).

[29] W. Kohn and L. J. Sham, Phys. Rev. **140**, A1133 (1965). https://doi.org/10.1103/PhysRev.140.A1133

[30] L. Lam and P. L. Platzman, Phys. Rev. B **9**, 5122 (1974). https://doi.org/10.1103/PhysRevB.9.5122

[31] P.-O. Lowdin, Phys. Rev. **97**, 1474 (1955). https://doi.org/10.1103/PhysRev.97.1474

[32] D. R. Hartree, Math. Proc. Camb. Phil. Soc. **24**(1), 89 (1928); https://doi.org/10.1017/S0305004100011919
Math. Proc. Camb. Phil. Soc. **24**(1), 111 (1928); https://doi.org/10.1017/S0305004100011920
Math. Proc. Camb. Phil. Soc. **24**(3), 426 (1928). https://doi.org/10.1017/S0305004100015954

[33] von V. Fock, Z. Phys. **61**, 126 (1930).

[34] J. C. Slater, Phys. Rev., **32**, 339 (1928); https://doi.org/10.1103/PhysRev.32.339 Phys.

Rev. **35**, 210 (1930). https://doi.org/10.1103/PhysRev.35.210

[35]T. von Koopmans, Physica (Amsterdam), **1**(1-6), 104 (1934). https://doi.org/10.1016/S0031-8914(34)90011-2

[36]H.-Y. Wang, Phys. Rev. B **62**(20), 13383 (2000). https://doi.org/10.1103/PhysRevB.62.13383

[37]H.-Y. Wang, *Green's Function in Condensed Matter Physics* (Alpha Science International ltd. and Science Press, Singapore, 2012). Chap. 13.